\RequirePackage{lineno}
\documentclass[aps,prd,twocolumn,superscriptaddress,showpacs,amsmath,amssymb]{revtex4}

\usepackage{comment}
\usepackage{multirow}
\usepackage{dcolumn}
\usepackage{bm}
\usepackage{setspace}
\usepackage{graphicx}
\includecomment{printrobustnotes}
\includecomment{printallnotes}
\usepackage{xspace}

\usepackage{dcolumn}
\usepackage{bm}

\newenvironment{cfigure1c}[1][tbp]{\begin{figure*}[#1]\centering}{\end{figure*}}

\newcommand{\note}[1]{\xspace}
\newcommand{\robustnote}[1]{\xspace}
\begin{printrobustnotes}
\renewcommand{\robustnote}[1]{\textbf{\textit{#1}}\xspace}
\end{printrobustnotes}
\begin{printallnotes}
\renewcommand{\note}[1]{\textbf{\textit{#1}}\xspace}
\renewcommand{\robustnote}[1]{\textbf{\textit{#1}}\xspace}
\end{printallnotes}
\newcommand{\ppbar}  {\ensuremath{p\bar{p}}\xspace}
\newcommand{\ttbar}  {\ensuremath{t\bar{t}}\xspace}

\newcommand{\Ht}     {\ensuremath{H_{T}}\xspace}
\newcommand{\et}     {\ensuremath{E_{T}}\xspace}
\newcommand{\etl}     {\ensuremath{E_{T}^{\mathrm{lepton}}}\xspace}
\newcommand{\etj}     {\ensuremath{E_{T}^{\mathrm{jet}}}\xspace}

\newcommand{\pt}     {\ensuremath{p_{T}}\xspace}

\newcommand{\met}    {\ensuremath{\text{\raisebox{.3ex}{$\not$}}\et}\xspace}

\newcommand{\dmt}{\ensuremath{\Delta{M}_{\text{top}}}\xspace}
\newcommand{\dmtr}{\ensuremath{\Delta m_t^{\text{reco}}}\xspace}
\newcommand{\dmttr}{\ensuremath{\Delta m_t^{\text{reco(2)}}}\xspace}

\newcommand{\chisq}  {\ensuremath{\chi^{2}}\xspace}

\newcommand{\mtop}    {\ensuremath{{M}_{\text{top}}}\xspace}
\newcommand{\mtopave} {\ensuremath{{M}_{\text{top}}^{\text{ave.}}\xspace}}

\newcommand{\genunit}[2]{\ensuremath{#1~\mathrm{#2}}\xspace}

\newcommand{\gev}[1]    {\genunit{#1}{GeV}}
\newcommand{\pb}[1]     {\genunit{#1}{pb}}

\newcommand{\gevc}[1]   {\ensuremath{#1~\mathrm{GeV}/c}}
\newcommand{\gevcc}[1]  {\ensuremath{#1~\mathrm{GeV}/c^{2}}}

\newcommand{\invfb}[1]  {\ensuremath{#1~\mathrm{fb}^{-1}}}

\newcommand{\gevccnoarg}{\ensuremath{\mathrm{GeV}/c^{2}}\xspace}

\newcommand{\measErr}[2]{\ensuremath{#1 \pm #2}\xspace}

\newcommand{\measStatSyst}[3]{\ensuremath{#1 \pm #2~(\textrm{stat}) \pm #3~(\textrm{syst})}\xspace}

\begin{document}
\title{Measurement of the Mass Difference Between Top and Anti-top Quarks}

\affiliation{Institute of Physics, Academia Sinica, Taipei, Taiwan 11529, Republic of China}
\affiliation{Argonne National Laboratory, Argonne, Illinois 60439, USA}
\affiliation{University of Athens, 157 71 Athens, Greece}
\affiliation{Institut de Fisica d'Altes Energies, ICREA, Universitat Autonoma de Barcelona, E-08193, Bellaterra (Barcelona), Spain}
\affiliation{Baylor University, Waco, Texas 76798, USA}
\affiliation{Istituto Nazionale di Fisica Nucleare Bologna, $^{ee}$University of Bologna, I-40127 Bologna, Italy}
\affiliation{University of California, Davis, Davis, California 95616, USA}
\affiliation{University of California, Los Angeles, Los Angeles, California 90024, USA}
\affiliation{Instituto de Fisica de Cantabria, CSIC-University of Cantabria, 39005 Santander, Spain}
\affiliation{Carnegie Mellon University, Pittsburgh, Pennsylvania 15213, USA}
\affiliation{Enrico Fermi Institute, University of Chicago, Chicago, Illinois 60637, USA}
\affiliation{Comenius University, 842 48 Bratislava, Slovakia; Institute of Experimental Physics, 040 01 Kosice, Slovakia}
\affiliation{Joint Institute for Nuclear Research, RU-141980 Dubna, Russia}
\affiliation{Duke University, Durham, North Carolina 27708, USA}
\affiliation{Fermi National Accelerator Laboratory, Batavia, Illinois 60510, USA}
\affiliation{University of Florida, Gainesville, Florida 32611, USA}
\affiliation{Laboratori Nazionali di Frascati, Istituto Nazionale di Fisica Nucleare, I-00044 Frascati, Italy}
\affiliation{University of Geneva, CH-1211 Geneva 4, Switzerland}
\affiliation{Glasgow University, Glasgow G12 8QQ, United Kingdom}
\affiliation{Harvard University, Cambridge, Massachusetts 02138, USA}
\affiliation{Division of High Energy Physics, Department of Physics, University of Helsinki and Helsinki Institute of Physics, FIN-00014, Helsinki, Finland}
\affiliation{University of Illinois, Urbana, Illinois 61801, USA}
\affiliation{The Johns Hopkins University, Baltimore, Maryland 21218, USA}
\affiliation{Institut f\"{u}r Experimentelle Kernphysik, Karlsruhe Institute of Technology, D-76131 Karlsruhe, Germany}
\affiliation{Center for High Energy Physics: Kyungpook National University, Daegu 702-701, Korea; Seoul National University, Seoul 151-742, Korea; Sungkyunkwan University, Suwon 440-746, Korea; Korea Institute of Science and Technology Information, Daejeon 305-806, Korea; Chonnam National University, Gwangju 500-757, Korea; Chonbuk National University, Jeonju 561-756, Korea; Ewha Womans University, Seoul, 120-750, Korea}
\affiliation{Ernest Orlando Lawrence Berkeley National Laboratory, Berkeley, California 94720, USA}
\affiliation{University of Liverpool, Liverpool L69 7ZE, United Kingdom}
\affiliation{University College London, London WC1E 6BT, United Kingdom}
\affiliation{Centro de Investigaciones Energeticas Medioambientales y Tecnologicas, E-28040 Madrid, Spain}
\affiliation{Massachusetts Institute of Technology, Cambridge, Massachusetts 02139, USA}
\affiliation{Institute of Particle Physics: McGill University, Montr\'{e}al, Qu\'{e}bec H3A~2T8, Canada; Simon Fraser University, Burnaby, British Columbia V5A~1S6, Canada; University of Toronto, Toronto, Ontario M5S~1A7, Canada; and TRIUMF, Vancouver, British Columbia V6T~2A3, Canada}
\affiliation{University of Michigan, Ann Arbor, Michigan 48109, USA}
\affiliation{Michigan State University, East Lansing, Michigan 48824, USA}
\affiliation{Institution for Theoretical and Experimental Physics, ITEP, Moscow 117259, Russia}
\affiliation{University of New Mexico, Albuquerque, New Mexico 87131, USA}
\affiliation{The Ohio State University, Columbus, Ohio 43210, USA}
\affiliation{Okayama University, Okayama 700-8530, Japan}
\affiliation{Osaka City University, Osaka 588, Japan}
\affiliation{University of Oxford, Oxford OX1 3RH, United Kingdom}
\affiliation{Istituto Nazionale di Fisica Nucleare, Sezione di Padova-Trento, $^{ff}$University of Padova, I-35131 Padova, Italy}
\affiliation{University of Pennsylvania, Philadelphia, Pennsylvania 19104, USA}
\affiliation{Istituto Nazionale di Fisica Nucleare Pisa, $^{gg}$University of Pisa, $^{hh}$University of Siena and $^{ii}$Scuola Normale Superiore, I-56127 Pisa, Italy, $^{mm}$INFN Pavia and University of Pavia, I-27100 Pavia, Italy}
\affiliation{University of Pittsburgh, Pittsburgh, Pennsylvania 15260, USA}
\affiliation{Purdue University, West Lafayette, Indiana 47907, USA}
\affiliation{University of Rochester, Rochester, New York 14627, USA}
\affiliation{The Rockefeller University, New York, New York 10065, USA}
\affiliation{Istituto Nazionale di Fisica Nucleare, Sezione di Roma 1, $^{jj}$Sapienza Universit\`{a} di Roma, I-00185 Roma, Italy}
\affiliation{Texas A\&M University, College Station, Texas 77843, USA}
\affiliation{Istituto Nazionale di Fisica Nucleare Trieste/Udine; $^{nn}$University of Trieste, I-34127 Trieste, Italy; $^{kk}$University of Udine, I-33100 Udine, Italy}
\affiliation{University of Tsukuba, Tsukuba, Ibaraki 305, Japan}
\affiliation{Tufts University, Medford, Massachusetts 02155, USA}
\affiliation{University of Virginia, Charlottesville, Virginia 22906, USA}
\affiliation{Waseda University, Tokyo 169, Japan}
\affiliation{Wayne State University, Detroit, Michigan 48201, USA}
\affiliation{University of Wisconsin, Madison, Wisconsin 53706, USA}
\affiliation{Yale University, New Haven, Connecticut 06520, USA}

\author{T.~Aaltonen}
\affiliation{Division of High Energy Physics, Department of Physics, University of Helsinki and Helsinki Institute of Physics, FIN-00014, Helsinki, Finland}
\author{S.~Amerio}
\affiliation{Istituto Nazionale di Fisica Nucleare, Sezione di Padova-Trento, $^{ff}$University of Padova, I-35131 Padova, Italy}
\author{D.~Amidei}
\affiliation{University of Michigan, Ann Arbor, Michigan 48109, USA}
\author{A.~Anastassov$^x$}
\affiliation{Fermi National Accelerator Laboratory, Batavia, Illinois 60510, USA}
\author{A.~Annovi}
\affiliation{Laboratori Nazionali di Frascati, Istituto Nazionale di Fisica Nucleare, I-00044 Frascati, Italy}
\author{J.~Antos}
\affiliation{Comenius University, 842 48 Bratislava, Slovakia; Institute of Experimental Physics, 040 01 Kosice, Slovakia}
\author{G.~Apollinari}
\affiliation{Fermi National Accelerator Laboratory, Batavia, Illinois 60510, USA}
\author{J.A.~Appel}
\affiliation{Fermi National Accelerator Laboratory, Batavia, Illinois 60510, USA}
\author{T.~Arisawa}
\affiliation{Waseda University, Tokyo 169, Japan}
\author{A.~Artikov}
\affiliation{Joint Institute for Nuclear Research, RU-141980 Dubna, Russia}
% ...

\collaboration{CDF Collaboration\footnote{With visitors from
$^a$Istituto Nazionale di Fisica Nucleare, Sezione di Cagliari, 09042 Monserrato (Cagliari), Italy,
$^b$University of California Irvine, Irvine, CA 92697, USA,
$^c$University of California Santa Barbara, Santa Barbara, CA 93106, USA,
$^d$University of California Santa Cruz, Santa Cruz, CA 95064, USA,
$^e$Institute of Physics, Academy of Sciences of the Czech Republic, 182~21, Czech Republic,
$^f$CERN, CH-1211 Geneva, Switzerland,
$^g$Cornell University, Ithaca, NY 14853, USA,
$^h$University of Cyprus, Nicosia CY-1678, Cyprus,
$^i$Office of Science, U.S. Department of Energy, Washington, DC 20585, USA,
$^j$University College Dublin, Dublin 4, Ireland,
$^k$ETH, 8092 Z\"{u}rich, Switzerland,
$^l$University of Fukui, Fukui City, Fukui Prefecture, Japan 910-0017,
$^m$Universidad Iberoamericana, Lomas de Santa Fe, M\'{e}xico, C.P. 01219, Distrito Federal,
$^n$University of Iowa, Iowa City, IA 52242, USA,
$^o$Kinki University, Higashi-Osaka City, Japan 577-8502,
$^p$Kansas State University, Manhattan, KS 66506, USA,
$^q$Brookhaven National Laboratory, Upton, NY 11973, USA,
$^r$University of Manchester, Manchester M13 9PL, United Kingdom,
$^s$Queen Mary, University of London, London, E1 4NS, United Kingdom,
$^t$University of Melbourne, Victoria 3010, Australia,
$^u$Muons, Inc., Batavia, IL 60510, USA,
$^v$Nagasaki Institute of Applied Science, Nagasaki 851-0193, Japan,
$^w$National Research Nuclear University, Moscow 115409, Russia,
$^x$Northwestern University, Evanston, IL 60208, USA,
$^y$University of Notre Dame, Notre Dame, IN 46556, USA,
$^z$Universidad de Oviedo, E-33007 Oviedo, Spain,
$^{aa}$CNRS-IN2P3, Paris, F-75205 France,
$^{bb}$Texas Tech University, Lubbock, TX 79609, USA,
$^{cc}$Universidad Tecnica Federico Santa Maria, 110v Valparaiso, Chile,
$^{dd}$Yarmouk University, Irbid 211-63, Jordan,
$^{ll}$Universite catholique de Louvain, 1348 Louvain-La-Neuve, Belgium,
$^{oo}$University of Z\"{u}rich, 8006 Z\"{u}rich, Switzerland,
$^{pp}$Massachusetts General Hospital and Harvard Medical School, Boston, MA 02114 USA,
$^{qq}$Hampton University, Hampton, VA 23668, USA,
$^{rr}$Los Alamos National Laboratory, Los Alamos, NM 87544, USA
}}
\noaffiliation

\date{\today}
\begin{abstract}
We present a measurement of the mass difference between top~($t$) and anti-top~($\bar{t}$) quarks using $\ttbar$
candidate events reconstructed in the final state with one lepton and multiple jets. We use the full data set of Tevatron $\sqrt{s} = 1.96$~TeV proton-antiproton 
collisions recorded by the CDF II detector, corresponding to an integrated luminosity of \invfb{8.7}. 
We estimate event-by-event the mass difference to construct templates for top pair signal events and 
background events. The resulting mass difference distribution in data compared to signal and background templates using a likelihood fit yields
$\dmt = {M}_{t} - {M}_{\bar{t}} = \gevcc{\measStatSyst{-1.95}{1.11}{0.59}}$ and is in agreement with the standard model prediction of no mass difference. 
\end{abstract}

\pacs{14.65.Ha, 13.85.Ni, 13.85.Qk, 12.15.Ff}
\maketitle

The laws of the standard model of particle physics~(SM) are invariant under the simultaneous transformations of charge conjugation, parity, and time reversal~({\it CPT}). 
Conservation of {\it CPT} is, therefore, fundamental and provides one of the most important constraints on the SM.
However, examining any possibility of {\it CPT} violation is important, as there are 
well-motivated extensions of the SM allowing for  {\it CPT} symmetry breaking~\cite{cptv}.  
In {\it CPT}-conserving models, particles and their anti-particles must have identical masses and widths. Thus,
any mass difference between a particle and its anti-particle would indicate a violation of {\it CPT}. 
{\it CPT} invariance has been tested for many elementary particles such as leptons and hadrons~\cite{cptexp,cptexp1}, but not in the bare quark except for the top quark~\cite{cpttop}. 
For all quarks except the top quark, direct mass measurements of bare quark are nearly impossible because the quark hadronization time scale is approximately an order of magnitude less than the quark decay time. 
After hadronization occurs, only the masses of hadrons are observable and give, at best, only an approximate estimate of the constituent quarks' masses. 
On the other hand, as the lifetime of the top quark is 
of the order of $10^{-24}$ seconds, it decays before hadronizing and a precision measurement of its mass and of the difference between the quark and anti-quark masses can be made.

Since the top quark discovery, close to three thousands of \ttbar candidate events have been collected per experiment at the Tevatron \ppbar collider. This sample makes measuring the top-quark mass~(\mtop) possible to an accuracy of approximately 0.5\%~(\mtop = \gevcc{173.2 \pm 0.9})~\cite{topmass_average} and the mass difference~($\dmt = {M}_{t} -{M}_{\bar{t}}$) between $t$ and $\bar{t}$ quarks to a comparable precision.  
The D0 collaboration performed several measurements of \dmt using matrix element analyses~\cite{d0cpt0,d0cpt1}. The most recent D0 result, based on a \invfb{3.6} data sample, reports  $\dmt = \gevcc{0.8 \pm 1.9}$, consistent with zero as predicted in the SM. The CDF collaboration performed a measurement using a \invfb{5.6} data sample~\cite{cdfcpt56} and found $\dmt = \gevcc{-3.3 \pm 1.7}$ which is also consistent with zero to within two standard deviations. To date, the most precise measurement is performed by the CMS collaboration, $\dmt=\gevcc{-0.44 \pm 0.53}$~\cite{cms5}. 

This paper reports on the final CDF measurement of \dmt 
based on the full Run~II data set 
corresponding to an integrated luminosity of \invfb{8.7}. We reconstruct the mass difference between $t$ and $\bar{t}$ quarks in each data event and compare its distribution with template distributions derived from Monte Carlo~(MC) model simulations to estimate \dmt. 
This is an update of a previous measurement that used a subset of the present data~\cite{cdfcpt56}. In addition to the larger data sample, we improve the jet energy calibration by applying an artificial neural network to achieve better jet energy resolution~\cite{nnjes}, as in a recent measurement of \mtop~\cite{cdfmass_lj}.
We also increase the size of the control samples and re-examine the systematic uncertainties.  

In the SM, $t$ and $\bar{t}$ quarks decay almost exclusively into a $W$ boson and a bottom quark ($t\rightarrow bW^{+}$ and $\bar{t}\rightarrow \bar{b}W^{-}$)~\cite{pdg}.
The case where one $W$ boson decays to a charged lepton~(electron or muon) and a neutrino ($W^{+}\rightarrow {\ell}^{+}\nu$ or $W^{-}\rightarrow {\ell}^{-}\bar{\nu}$ including the cascade decay of $W \rightarrow \tau\nu$ and $\tau \rightarrow \ell \nu$) and 
the other to a pair of jets, defines the lepton+jets channel. 
To select $\ttbar$ candidate events in this channel, we require one electron (muon) with 
$\et > \gev{20}$ ($\pt > \gevc{20}$) and pseudorapidity $|\eta|<1.1$~\cite{cdfco}.  
We also require large missing transverse energy~\cite{met} ($\met > \gev{20}$) and at least four jets. 
Jets are reconstructed applying a cone algorithm with radius 
$\Delta R = \sqrt{(\Delta \eta)^2+(\Delta \phi)^2} = 0.4$~\cite{jetclu}.
Besides the standard jet energy scale corrections~\cite{jes}, we use an artificial neural network that includes additional information to the calorimeter one, such as jet momentum from the charged particles inside the jet~\cite{nnjes}. This additional information improves the resolution on the reconstructed jet variables, resulting in approximately a 10\% improvement in statistical precision.  
Jets originating from 
$b$ quarks are identified~(tagged) using a secondary vertex tagging algorithm~\cite{secvtx}. 
In order to optimize the background reduction and to improve the statistical power of the measurement, we divide the sample of \ttbar candidates into subsamples with zero~(0-tag), one~(1-tag), and two or more~(2-tag) $b$-tagged jets. 

For the 0-tag events, we require exactly four {\it tight} jets~(transverse energy $\et > \gev{20}$ and $|\eta|<2.0$). In case of the 1-tag and 2-tag events, three tight jets and one or more {\it loose} jets ($\et > \gev{12}$ and $|\eta| <2.4$) are required. 
To reduce background contributions to the 0-tag or 1-tag samples, we require the scalar sum of transverse energies in the event, 
$\Ht = \etl + \met + \sum_{\text{four jets}}\etj$, to exceed 250~GeV. The $\Ht$ requirement is not applied to the 2-tag events because of the small background contribution in this subsample. We divide the 1-tag and 2-tag samples into subsamples based on the number of tight jets. We denote as  {\it tight subsample} the sample requiring exactly four tight jets and {\it loose subsample} the sample consisting of the remaining events. This results in five subsamples: 0-tag, 1-tagL, 1-tagT, 2-tagL, and 2-tagT, where T and L denote tight and loose subsamples, respectively.  

\begin{table*}[ht]
\begin{center}
\caption[Expected number of background and signal events]{
Expected and observed numbers of signal and background events 
assuming a \ttbar production cross section $\sigma_{\ttbar}$ = \pb{7.45} and \mtop = \gevcc{172.5}. 
}
\begin{ruledtabular}
\label{table_background}
\begin{tabular}{lccccc}
             &0-tag       & 1-tagL     &1-tagT       &2-tagL      & 2-tagT\\
\hline
$W$+jets       & 778$\pm$219& 197$\pm$69 & 114$\pm$42  &11.4$\pm$4.9& 8.0$\pm$3.4\\
$Z$+jets       &55.7$\pm$4.9&10.3$\pm$1.2& 6.7$\pm$0.8 & 0.8$\pm$0.2& 0.5$\pm$0.1\\
Single top   & 5.1$\pm$0.4&11.7$\pm$1.0& 7.2$\pm$0.6 & 2.2$\pm$0.2& 1.7$\pm$0.2\\
Diboson      &63.9$\pm$5.9&11.7$\pm$1.5& 9.0$\pm$1.2 & 0.9$\pm$0.2& 0.9$\pm$0.2\\
QCD multijet & 133$\pm$107&31.7$\pm$1.2&20.9$\pm$16.9& 4.3$\pm$4.3& 2.9$\pm$3.5\\
\hline
Total background   &1038$\pm$244& 262$\pm$70 & 158$\pm$45  &19.5$\pm$6.5&14.0$\pm$5.0\\
\ttbar signal& 620$\pm$83 & 694$\pm$87 & 847$\pm$105 & 188$\pm$29 & 294$\pm$45\\
\hline
Expected     &1658$\pm$257& 957$\pm$111&1005$\pm$114 & 208$\pm$30 & 308$\pm$45\\
Observed     &1712        & 919        & 1018        & 214        &286\\
\end{tabular}
\end{ruledtabular}
\end{center}
\end{table*}

The primary sources of background contributions  are $W$+jets and QCD multijet processes. 
To estimate the contribution of each process, we use a combination of data- and MC-based techniques described in Ref.~\cite{xsection,tmtold}. For the $Z$+jets, diboson, single top quark, and \ttbar events
we normalize the number of simulated events using their theoretical cross sections~\cite{smcross1,smcross2,thcross}. We use the data-driven techniques described in Ref.~\cite{qcdback} to estimate the QCD multijet background. 
The $W$+jets background shape is modeled using MC generated samples but the number of events is derived from the data sample by subtracting all other contributions, including the \ttbar signal, from the data events. 
Table~\ref{table_background} summarizes the data sample composition.  The distribution of $\Ht$ is shown in Fig.~\ref{ref:valid} for data with 0-tag and one or more $b$-tag~(Tagged) with the predictions from our signal and background models. 

\begin{cfigure1c}
\begin{tabular}{cc}
\includegraphics[width=0.43\textwidth]{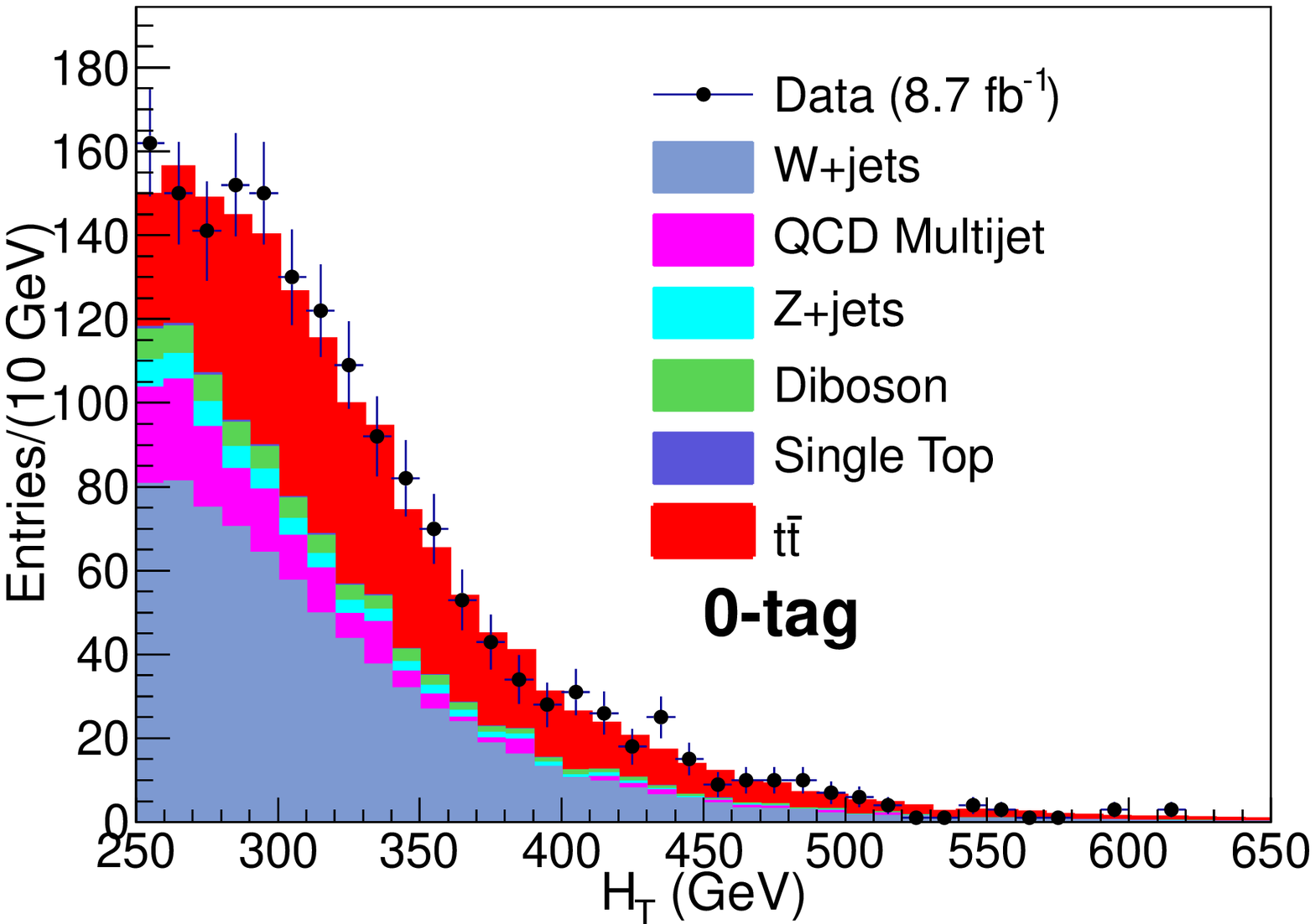} &
\includegraphics[width=0.43\textwidth]{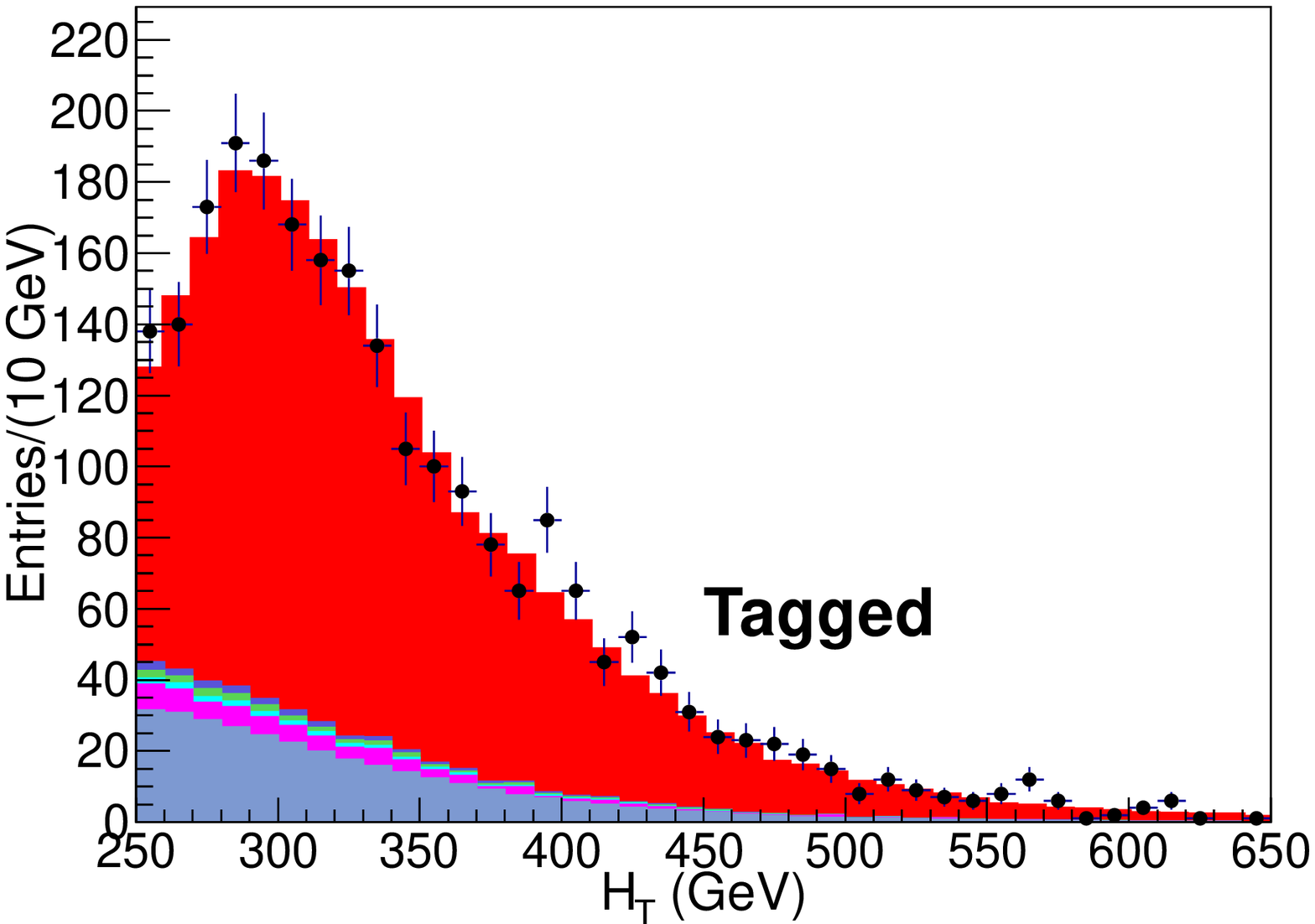} \\
\end{tabular}
\caption[Validation]{$\Ht$ distribution for zero 
$b$-tagged~(0-tag) events and one or more $b$-tagged~(Tagged)  events.
}
\label{ref:valid}
\end{cfigure1c}

We assume that all selected events are lepton+jets \ttbar events and reconstruct 
\dmt, event-by-event, using a special-purpose kinematic fitter~\cite{cdfcpt56}.
Measured four-vectors of the lepton and jets are corrected for known effects as described in Ref.~\cite{jes}, 
and appropriate resolutions are assigned. The unclustered transverse energy~($U_{T}$) is estimated as a sum of all transverse energy in the calorimeters that is not associated with the primary lepton or with one of the leading four jets. It is used to calculate the neutrino transverse momentum. The longitudinal momentum of the neutrino is a free parameter which is effectively determined by the constraint on the invariant mass of the leptonically-decaying $W$ boson. 
To estimate \dmt, we define a kinematic $\chi^{2}$ function, 
\begin{linenomath*}
\begin{eqnarray}
\label{eq_chi2}
\chisq & = &
\Sigma_{i = \ell, 4 \text{jets}} {(p_T^{i,\text{fit}} - p_T^{i,\text{meas}})^2 / \sigma_i^2} \nonumber \\
& + & \Sigma_{k = x,y} {(U_{T_k}^{\text{fit}} - U_{T_k}^{\text{meas}})^2 / \sigma_k^2} \nonumber \\
& + & {({M}_{jj} - {M}_W)^2 / \Gamma_W^2} 
 + {({M}_{\ell\nu} - {M}_W)^2 / \Gamma_W^2} \nonumber \\
 & + &  {\{{M}_{bjj} - (\mtopave+{dm}_{\text{reco}}/2)\}^2 / \Gamma_t^2} \nonumber \\
 & + & {\{{M}_{b\ell\nu} - (\mtopave-{dm}_{\text{reco}}/2)\}^2 / \Gamma_t^2},
\end{eqnarray}
\end{linenomath*}
where $dm_{\text{reco}}$ is obtained at the lowest $\chi^2$ and represents 
the reconstructed mass difference between the hadronically- and leptonically-decaying top quarks, ${M}_{bjj}-{M}_{b\ell\nu}$. 
In Eq.~(1), we constrain the lepton \pt and 
the four leading jets \pt to their measured 
values and uncertainties~($\sigma_i$). We also constrain 
$U_{T}$ in the second term of Eq.~\ref{eq_chi2}. In the remaining terms, we constrain the $W$ boson mass~($M_{W}$) to ${M}_W$=\gevcc{80.4}~\cite{wmass} and the average of $t$ and $\bar{t}$ masses to $\mtopave$=\gevcc{172.5}. 
The quantities ${M}_{jj}, {M}_{\ell\nu}, {M}_{bjj}$, and ${M}_{b\ell\nu}$ 
refer to the invariant masses of the particles denoted in the subscripts. 
The total widths of the $W$ boson, $\Gamma_W$ = \gev{2.1}, and of the top quark, $\Gamma_t$ = \gev{1.5}, are taken from Ref.~\cite{pdg}. We assume that the total widths of the $t$ and $\bar t$ quarks are equal.  
Determining the reconstructed mass difference of $t$ and $\bar{t}$, $\dmtr$, requires the
identification of the particle type ($t$ or $\bar t$), which is achieved using the electric charge
of the lepton~($Q_{\text{lepton}}$),
$\dmtr = -Q_{\text{lepton}} \cdot {dm}_{\text{reco}}$.
In the events with a positive~(negative) lepton, $t$~($\bar{t}$) decays leptonically and $\bar{t}$~($t$) decays hadronically. Because of the different resolutions of the jets, lepton, and unclustered energy, the distribution of reconstructed mass from the hadronic top quark is different with that of the leptonic top quark. 
To improve the resolution of the \dmtr, and allow using the appropriate distribution in the hadronic-to-leptonic and in the leptonic-to-hadronic mass difference, we divide each subsample into the two new subsamples based on the lepton charge. We then have ten subsamples in total. 

Assuming that the leading four jets in any event come from the four final 
quarks of the \ttbar lepton+jets decay at the hard scattering level, there are 12, 6, and 2 possible jet-to-quark assignments for 0-tag, 1-tag, and 2-tag samples, respectively. 
The $\chi^2$ minimization is performed for each jet-to-quark assignment, and \dmtr is taken from the assignment 
that yields the lowest $\chi^2$ ($\chi^2_{\text{min}}$). The $b$-tagged~(zero $b$-tag) events with 
$\chi^2_{\text{min}} > 9.0$~($\chi^2_{\text{min}} > 3.0$) 
are rejected due to the poorly reconstructed kinematical properties. 
To increase the statistical power of the measurement, we employ an 
additional observable, \dmttr, which corresponds to the 2nd lowest $\chi^2$ in the jet-to-quark combinatorics. Although it has a poorer sensitivity, \dmttr 
provides additional information on \dmt and reduces the statistical 
uncertainty by approximately 10\%. We use two observables~(\dmtr and \dmttr) simultaneously for the measurement.

Using {\sc madgraph}~\cite{madgraph}, we generate \ttbar signal samples
with \dmt between \gevcc{-20} and \gevcc{20} in \gevcc{2} intervals.  
Parton showering of the signal events is simulated with 
{\sc pythia}~\cite{pythia}, and the CDF II detector is simulated using a {\sc geant}-based 
software package~\cite{cdfsim}. 

We estimate the probability density functions~(PDFs) of signal and
background using the kernel density estimation~\cite{KDEHEP,tmt19}.
We construct the two dimensional PDFs that account for the correlation between \dmtr and \dmttr.
First, at discrete values of \dmt from \gevcc{-20} to \gevcc{20}, we estimate 
the PDFs for the observables from the above-mentioned {\sc madgraph} $\ttbar$ samples. 
We interpolate the MC distributions 
to find PDFs for arbitrary values of \dmt using the local polynomial smoothing
method~\cite{lps}.  Then, we fit the signal and background PDFs to the unbinned distributions 
observed in the data using a maximum likelihood fit~\cite{eml}. 
Separate likelihoods are built for the ten subsamples, and the overall likelihood is obtained by multiplying them together.
References~\cite{cdfmass_lj,tmt19} provide detailed information about this technique. 

We calibrate the method using the fully simulated MC experiments. We perform 3000 simulated experiments
for each of eleven equally-spaced $\dmt$ values ranging from \gevcc{-10} to \gevcc{10}. 
The fit estimates and their uncertainties in the simulated experiments are found to be unbiased. 

\begin{table}
\begin{center}
\caption[Summary of systematics]{Summary of systematic uncertainties on \dmt.}
\begin{ruledtabular}
\label{systtablesummary}
\begin{tabular}{lc}
Source              & Uncertainty~(\gevccnoarg) \\
\hline
Signal modeling                    & 0.14 \\
Parton showering                   & 0.17 \\
$b$ and $\bar{b}$ jets asymmetry   & 0.38 \\
Higher-Order effect                & 0.16 \\
Jet energy scale                   & 0.07 \\
Parton distribution functions      & 0.12 \\
$b$-jet energy scale               & 0.05 \\
Background shape                   & 0.20 \\
Gluon fusion fraction              & 0.05 \\
Initial and final state radiation  & 0.10 \\
Finite Monte Carlo samples         & 0.07 \\
Lepton energy scale                & 0.06 \\
Multiple hadron interaction        & 0.05 \\
Color reconnection                 & 0.23 \\
\hline
Total systematic uncertainty       & 0.59 \\
\end{tabular}
\end{ruledtabular}
\end{center}
\end{table}

We examine a variety of systematic effects that could affect the \dmt measurement. To estimate the systematic uncertainties,
we compare the results from simulated experiments in which we vary relevant
parameters within one standard deviation. 
We estimate the systematic uncertainties in the assumptions of \mtop = \gevcc{172.5} and \dmt = \gevcc{0.0}.  
All systematic uncertainties are summarized 
in Table~\ref{systtablesummary}.   
The dominant source of systematic uncertainty is attributed to a possible difference in the detector response
between $b$ and $\bar{b}$ jets.
To estimate this effect, we select a $b\bar{b}$ sample by requiring exactly two $b$-tagged jets per event using a sample triggered on jet~($E_{T}>\gev{20}$). In addition, one $b$-tagged jet is required to contain a soft muon from leptonic decay so that the charge tendency of the $b$ quark associated with the jet can be estimated.  
The energy scale of $b$ and $\bar{b}$ influenced jet events in data is compared with di-jet MC events in which 
we estimate the \pt imbalance~(\pt of $b$ influenced jets minus \pt of $\bar{b}$ influenced jets devided by average \pt) difference between the data and the MC events and obtain $-0.44\pm0.40\%$. To calculate the \pt imbalance difference from $b$ and $\bar{b}$ jets, we estimate the fraction of the $b$ quark flavors associated with same charge of the soft muons. We obtain the \pt imbalance difference to be $-0.73\pm0.67\%$ with considering incorrect charge events anti-correlately. 
We perform simulated experiments by varying the $b$ and $\bar{b}$ energy within their \pt imbalance difference. 
The possible difference of calorimeter responses between $c$ and $\bar{c}$ jets can be a source of systematic uncertainty. With an assumption of same asymmetry between $b$ and $\bar{b}$ jets as $c$ and $\bar{c}$ jets, we obtain a tiny uncertainty, \gevcc{0.03}, which is neglected. 
We estimate the signal modeling uncertainty by using simulated experiments with events generated with {\sc madgraph} and {\sc pythia}. 
We also estimate a parton showering uncertainty by applying different showering models 
({\sc pythia} and {\sc herwig}~\cite{herwig}) to a sample generated with {\sc alpgen}~\cite{alpgen}. 
Higher-order effects are estimated using {\sc mc@nlo} generator~\cite{mcnlo}.
The background shape systematic uncertainty accounts for the variation of the background composition as well as the overall background fraction. We also consider changes in the shapes by varying the $Q^{2}$ used in the calculation of hard scattering and showering. 
The color reconnection systematic uncertainty~\cite{CR} is evaluated using the samples with and without color reconnection effects in {\sc pythia} tunes~\cite{CR_tune}. We use two samples with angular ordering for jet showers~(tune $A-Pro$ and tune $ACR-Pro$), same as the nominal samples of \dmt measurement.  We also have a cross check using the other two samples with $\pt$ ordering for jet showers and new underlying-event model~($Perugia 0$ and $Perugia NOCR$) and find a similar uncertainty.  
We vary the parameters of parton distribution functions to account for systematic effects. 
The jet energy scale uncertainty, the dominant uncertainty in most of the \mtop measurements, is partially canceled in the $t$ and $\bar{t}$ mass difference. 
Other sources of systematic effects, 
including uncertainties in gluon radiation, multiple hadron interaction, finite size of MC samples, $b$-jet energy scale,
and lepton energy scale, give small contributions. Because we assume the average \mtop to be \gevcc{172.5}, the \mtop dependence can be a possible source of systematic uncertainty.  We perform the simulated experiments using different \ttbar signal samples of \mtop from \gevcc{170.0} to \gevcc{175.0} with \gevcc{0.5} steps. All samples have \dmt = \gevcc{0}. We find the measured \dmt values, \gevcc{0.01\pm0.08} in the fit, are consistent with zero. 
The total systematic uncertainty of \gevcc{0.59} is calculated as a quadrature sum of the listed uncertainties. The details of systematic uncertainty evaluations are in Ref~\cite{topmass_average,tmtold,tmt19}. 
 
\begin{cfigure1c}
\begin{tabular}{cc}
\includegraphics[width=0.43\textwidth]{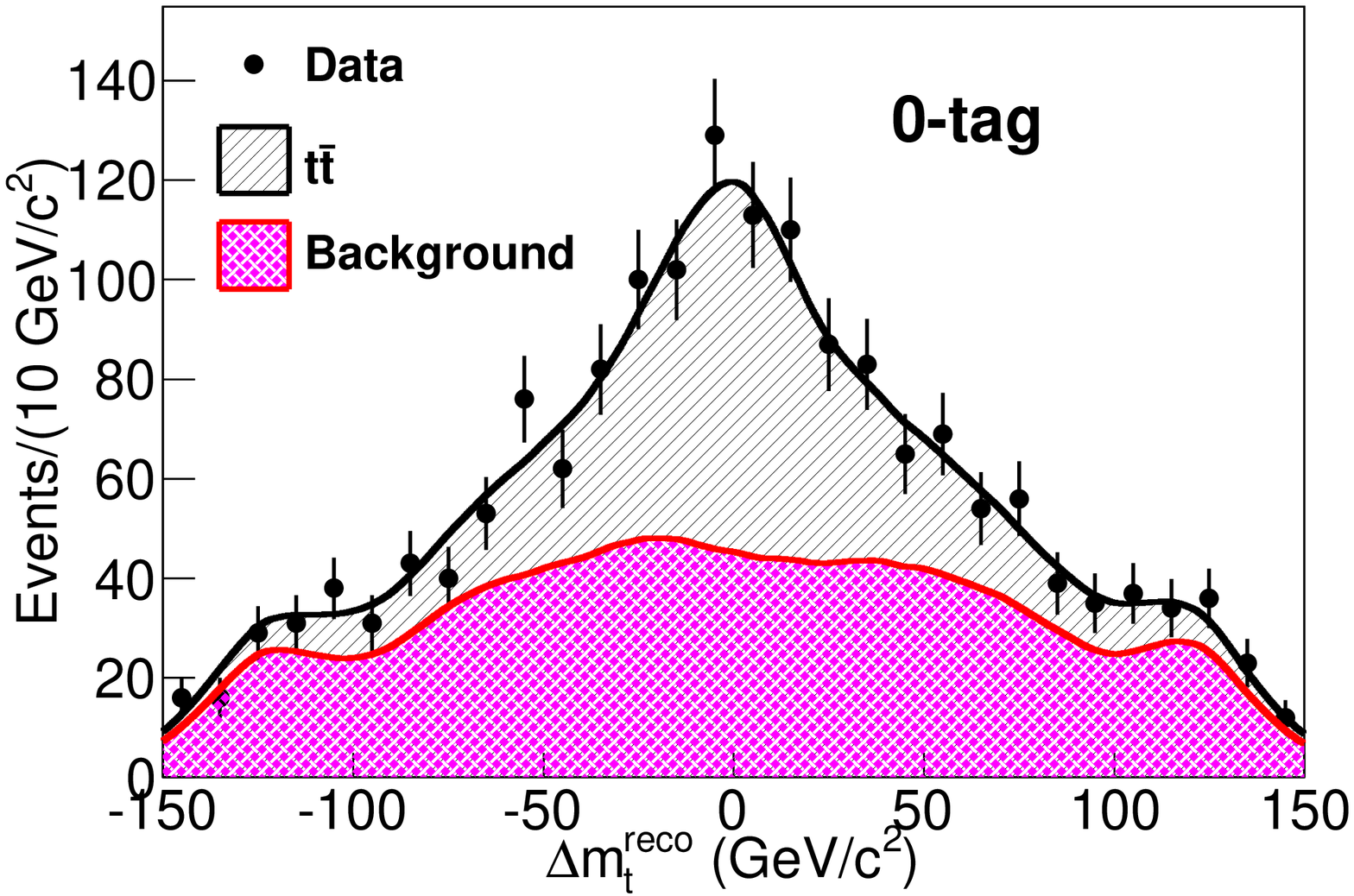} &
\includegraphics[width=0.43\textwidth]{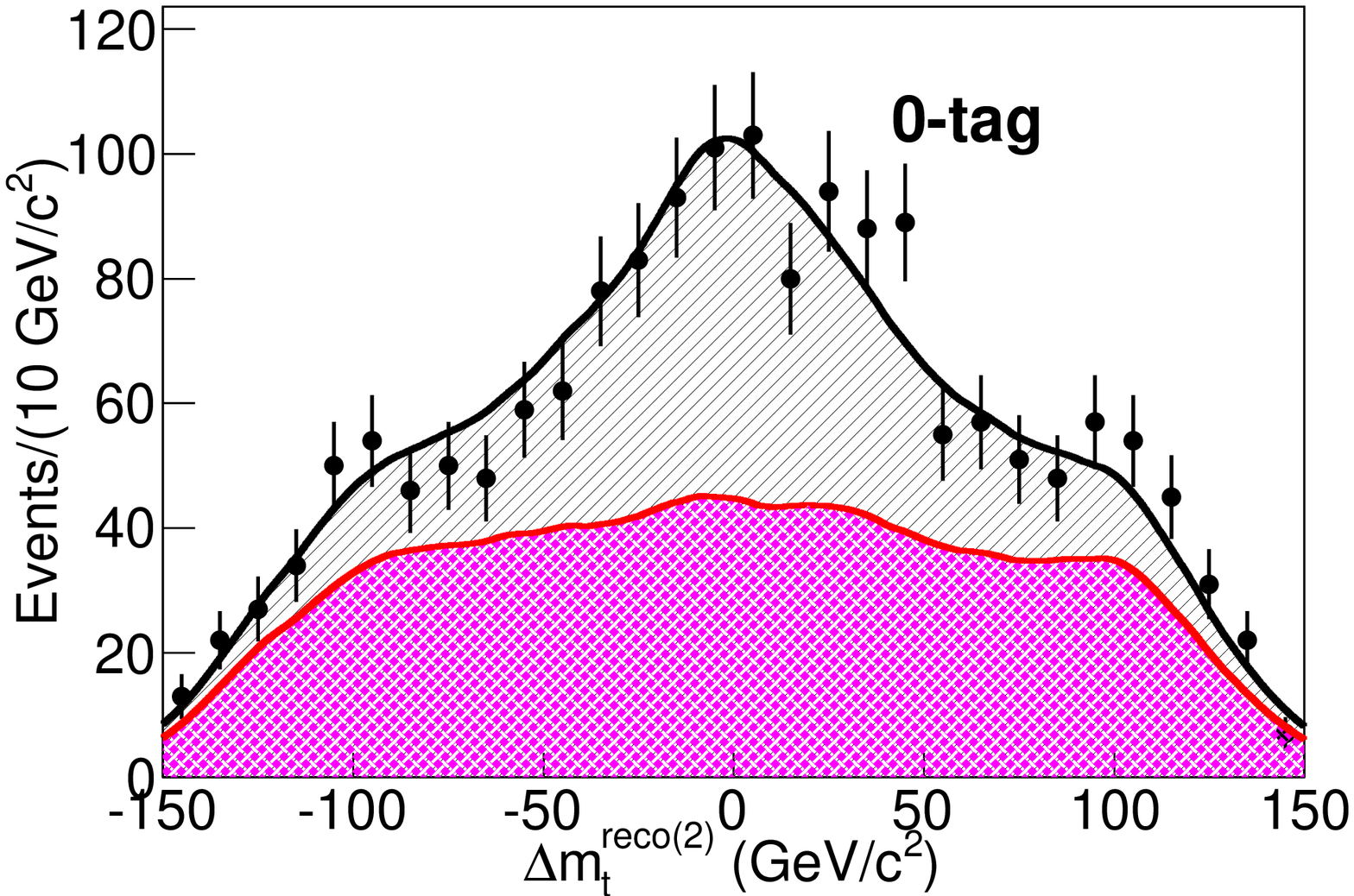} \\
\includegraphics[width=0.43\textwidth]{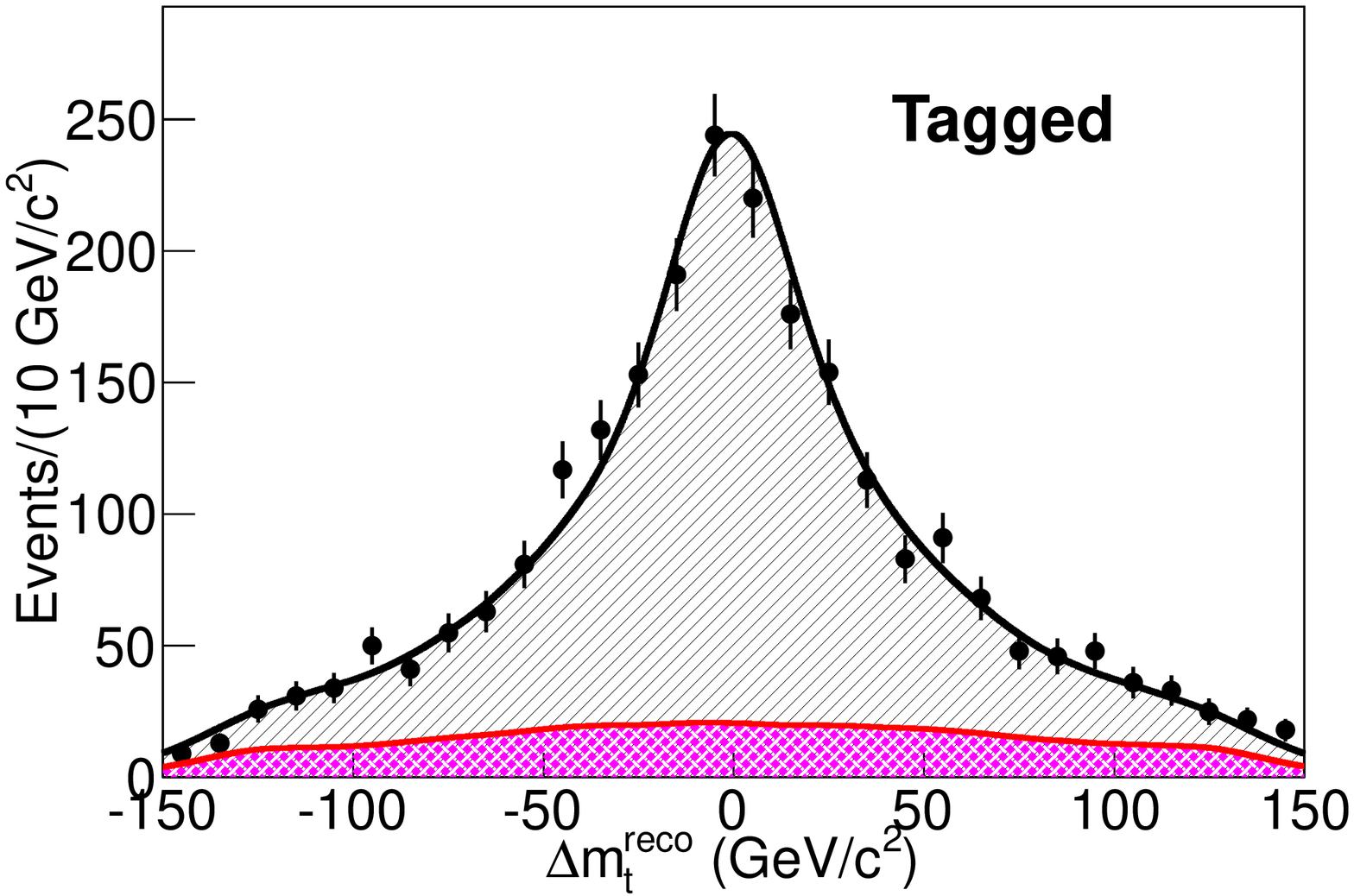} &
\includegraphics[width=0.43\textwidth]{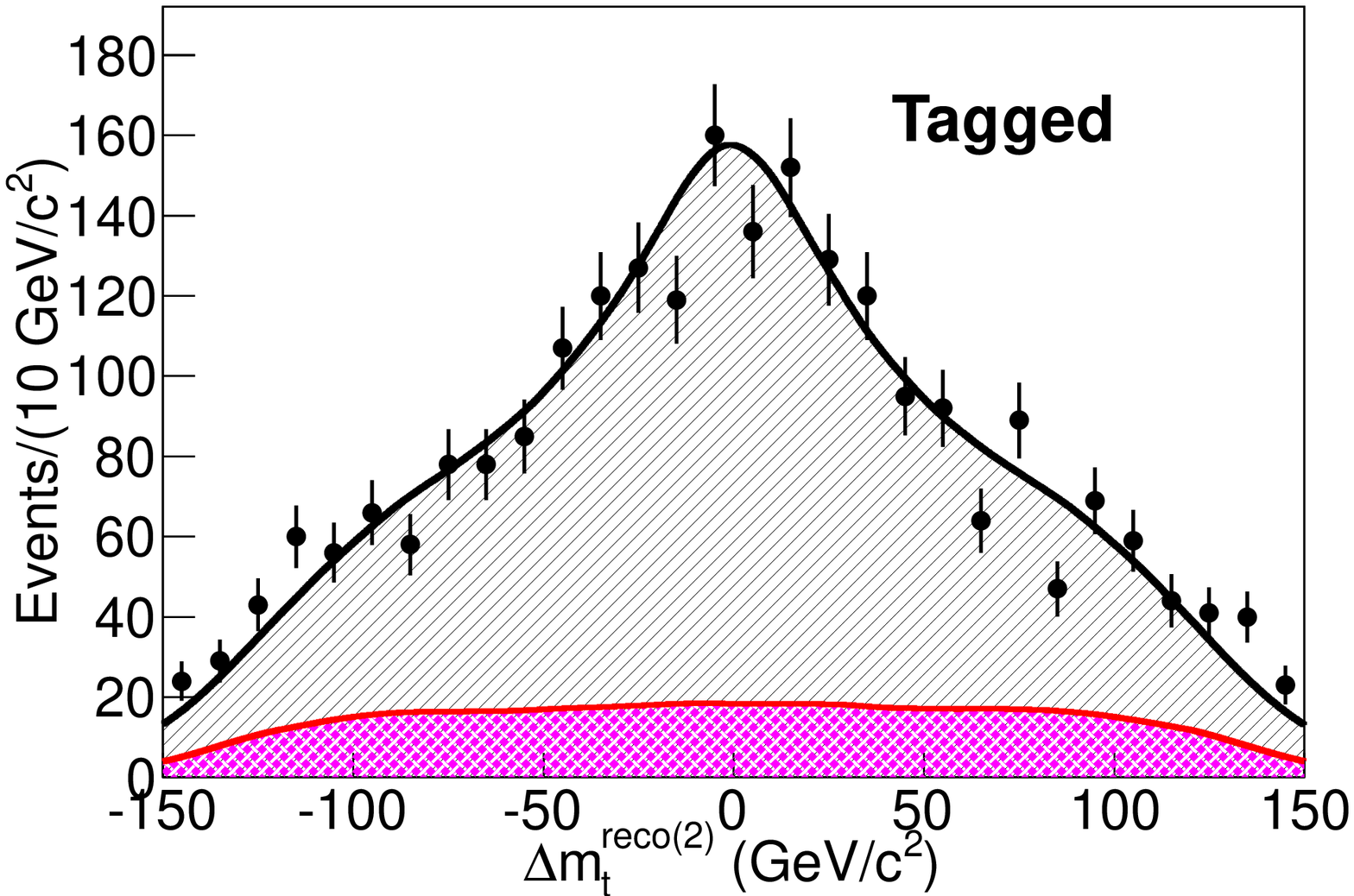} \\
\end{tabular}
\caption[Data fit]{Distributions of $\dmtr$ and $\dmttr$ used to extract $\dmt$ for zero 
$b$-tagged~(0-tag) events and one or more $b$-tagged~(Tagged)  events. 
The data are overlaid with predictions from the kernel density estimation probability 
distributions assuming \dmt~=~\gevcc{0}. The fitted number of signal and background events are used. 
}
\label{ref:fit}
\end{cfigure1c}

The resulting mass difference is
$$\dmt  = \gevcc{\measStatSyst{-1.95}{1.11}{0.59}}.$$
Figure~\ref{ref:fit} shows the observed distributions of the observables used for 
the \dmt measurement. The density estimates for \ttbar signal events with \dmt = \gevcc{0} and for background events are overlaid.  

In conclusion, we examine the mass difference between $t$ and $\bar{t}$ quarks in 
the lepton+jets channel using CDF II data corresponding to an integrated luminosity of \invfb{8.7} from \ppbar 
collisions at $\sqrt{s}$~=~1.96~TeV. We measure the mass difference 
to be $\dmt = {M}_{t} - {M}_{\bar{t}} = \gevcc{\measStatSyst{-1.95}{1.11}{0.59}} = \gevcc{\measErr{-1.95}{1.26}}$. 
This result is consistent with $\dmt = \gevcc{0}$ and conservation of {\it CPT} symmetry. 

\begin{acknowledgments}
	We thank the Fermilab staff and the technical staffs of the participating institutions for their vital contributions. This work was supported by the U.S. Department of Energy and National Science Foundation; the Italian Istituto Nazionale di Fisica Nucleare; the Ministry of Education, Culture, Sports, Science and Technology of Japan; the Natural Sciences and Engineering Research Council of Canada; the National Science Council of the Republic of China; the Swiss National Science Foundation; the A.P. Sloan Foundation; the Bundesministerium f\"ur Bildung und Forschung, Germany; the Korean World Class University Program, the National Research Foundation of Korea; the Science and Technology Facilities Council and the Royal Society, UK; the Russian Foundation for Basic Research; the Ministerio de Ciencia e Innovaci\'{o}n, and Programa Consolider-Ingenio 2010, Spain; the Slovak R\&D Agency; the Academy of Finland; and the Australian Research Council (ARC). 
\end{acknowledgments}


\begin{thebibliography}{99}
\bibitem{cdfco} We use a right-handed spherical coordinate system with the origin at the center of the detector. $\theta$ and $\phi$ are the polar and azimuthal angles, respectively.  The pseudo-rapidity is defined by $\eta = -\ln \tan(\theta/2)$.  The transverse momentum and energy are defined by $\pt = p \sin(\theta)$ and $\et = E \sin(\theta)$, respectively, where $p$ and $E$ are the momentum and energy of the particle.
\bibitem{met} The missing transverse energy, an imbalance of energy in the transverse plane of the detector, is defined by 
$\met = -\vert \sum_{towers}\et{\hat n_{T}} \vert$, where ${\hat n_{T}}$ is the unit vector normal to the beam and pointing to a given calorimeter tower and $\et$ is the transverse energy measured in that tower.
\end{thebibliography}
\end{document}